\documentclass{aa}
\usepackage{graphics,epsfig,lscape,times}

\usepackage{natbib}
\bibpunct{(}{)}{;}{a}{}{,}

\newcommand{\ergcm}[1]{$\times 10^{#1}$ erg cm$^{-2}$ s$^{-1}$}
\newcommand{\oergcm}[1]{$10^{#1}$ erg cm$^{-2}$ s$^{-1}$}
\newcommand{\ergs}[1]{$\times 10^{#1}$ erg s$^{-1}$}
\newcommand{\oergs}[1]{$10^{#1}$ erg s$^{-1}$}
\newcommand{\hcm}[1]{$\times 10^{#1}$ cm$^{-2}$}
\newcommand{\ohcm}[1]{$10^{#1}$ cm$^{-2}$}
\newcommand{\expo}[1]{$\times 10^{#1}$}
\newcommand{\oexpo}[1]{$10^{#1}$}
\newcommand{\nh}{N$_{\rm H}$}

\newcommand{\ct}{cts s$^{-1}$}

\newcommand{\HII}{\ion{H}{II}}
\newcommand{\Hone}{\ion{H}{I}}
\newcommand{\Hmol}{H$_2$}
\newcommand{\SII}{\ion{S}{II}}
\newcommand{\Halp}{H${\alpha}$}
\newcommand{\exo}{\hbox{\object{EXO\,053109-6609.2}}}
\newcommand{\rxbe}{\hbox{\object{RX\,J0529.8-6556}}}
\newcommand{\rxob}{\hbox{\object{RX\,J0532.5-6551}}}
\newcommand{\xmmbex}{\hbox{\object{XMMU\,J053011.2-655122}}}
\newcommand{\xmmsnr}{\hbox{\object{XMMU\,J053226.2-655352}}}
\newcommand{\xmmsss}{\hbox{\object{XMMU\,J053056.2-654809}}}
\begin{document}
 
\title{Deep XMM-Newton observation of a northern LMC field: \\
       I. Selected X-ray sources
        \thanks{XMM-Newton is an ESA Science Mission
               with instruments and contributions directly funded by ESA Member
               states and the USA (NASA)}
}
 
\author{F.~Haberl \and K.~Dennerl \and W.~Pietsch}

\titlerunning{Deep XMM-Newton observation of a northern LMC field}
\authorrunning{Haberl et al.}
 
\offprints{F. Haberl, \email{fwh@mpe.mpg.de}}
 
\institute{Max-Planck-Institut f\"ur extraterrestrische Physik,
           Giessenbachstra{\ss}e, 85748 Garching, Germany}
 
\date{Received; accepted}
 
\abstract{First results from a deep XMM-Newton observation of a field 
in the Large Magellanic Cloud (LMC) near the northern rim of the supergiant shell 
LMC\,4 are presented. Spectral and temporal analyses of a sample of selected
X-ray sources yielded two new candidates for supernova remnants, a supersoft
X-ray source and a likely high mass X-ray binary (HMXB) pulsar. From the fourteen brightest
sources up to ten are active galactic nuclei in the background of the galaxy 
which can be used as probes for the interstellar medium in the LMC. From the three 
previously known HMXBs the Be/X-ray binary \exo\ was the brightest source in the field,
allowing a more detailed analysis of its X-ray spectrum and pulse profile. During the pulse 
\exo\ shows eclipses of the X-ray emitting areas with increased photo-electric 
absorption before and after the eclipse. The detection of X-ray pulsations with a period 
of 69.2 s is confirmed
for \rxbe\ and a possible period of 272 s is discovered from \xmmbex.
The results are discussed with respect to individual sources as well 
as in the view of source population studies in the vicinity of the supergiant 
shell LMC\,4.
}
 
\maketitle
 
\section{Introduction}

Due to their proximity the Large and Small Magellanic Cloud (LMC, SMC) were always 
subject to X-ray surveys for satellites carrying imaging instruments. The most 
sensitive and most 
complete survey is available from ROSAT data obtained between 1990 and 1998 in 
the energy range 0.1--2.4 keV. In particular pointed observations with the
PSPC detector with its large field of view covered in total $\sim$59 and $\sim$18 square 
degrees of the LMC and SMC regions on the sky 
\citep{1999A&AS..139..277H,2000A&AS..142...41H}.
The ROSAT PSPC and HRI observations revealed about 1000 and 750 X-ray sources in 
the direction of the LMC and SMC, respectively 
\citep[see also][]{2000A&AS..143..391S,2000A&AS..147...75S}.

The large number of X-ray sources in the Magellanic Cloud fields allows  
population studies on rich samples of various kinds of source types powered
by different mechanisms. An unusually high concentration of Be/X-ray binary 
systems was found in the SMC 
\citep{2000A&A...359..573H}. Detectors sensitive to higher energies on board ASCA
and XTE detected pulsations from many new Be/X-ray transients during outburst
\citep{2002Yokogawa}. The high sensitivity of the XMM-Newton instruments promises
the detection of pulsations to much lower X-ray fluxes \citep{2001A&A...369L..29S}
usually observed from Be/X-ray binaries in their low state. 

To study the source population in the LMC to flux limits below \oergcm{-14} a deep
$\sim$60 ks XMM-Newton observation was performed as part of the telescope scientist
guaranteed time. The selected field, aimed at 
RA = 05$^{\rm h}$31$^{\rm m}$20$^{\rm s}$ and Dec = --65\degr57\arcmin38\arcsec,
was chosen because of its location on the rim of the supergiant shell (SGS) LMC\,4
\citep{1980MNRAS.192..365M} and the 
simultaneous coverage of three known high mass X-ray binaries 
(HMXBs). These are : 1) The Be/X-ray binary pulsar \exo\ 
discovered in outburst during EXOSAT observations of the LMC\,X-4 region in 1983 
\citep{1985SSRv...40..379P}. X-ray pulsations with a pulse period of 13.7 s were 
detected for the first time in ROSAT data by \citet{1996rftu.proc..131D}.
2) An outburst observed from \rxbe\ by ROSAT and the detection 
of 69 s X-ray pulsations led to the suggestion by \citet{1997A&A...318..490H} that 
\rxbe\ is also a Be/X-ray binary. Optical spectroscopy by \citet{2002A&A...385..517N}
confirmed this. 3) The highly variable source \rxob\ proposed as first HMXB in the LMC 
powered by accretion from the wind of an OB supergiant companion \citep{1995A&A...303L..49H} 
which is supported by optical spectroscopy \citep{2002A&A...385..517N}.

Active galactic nuclei (AGN) behind nearby galaxies are of interest for two main
reasons. They can be used to define a precise reference coordinate system 
\citep[e.g.][]{2000AJ....120..845A} and provide
line of sight probes to study the interstellar medium in the galaxies
\citep[e.g.][]{2001A&A...371..816K,2001A&A...365L.208H}.
The new X-ray observatories Chandra and XMM-Newton will largely increase the number
of background AGN as first results from LMC observations demonstrate 
\citep{2001A&A...365L.208H,2002ApJ...569L..15D}.

In this article first results from the deep XMM-Newton observation are presented.
The paper concentrates on spectral and temporal analyses of a selected sample of
X-ray sources which includes the three known HMXBs and new candidates for supernova
remnants (SNRs), a supersoft source (SSS), a likely Be/X-ray binary pulsar and ten
candidates for background AGN.

\begin{figure*}
\resizebox{\hsize}{!}{\includegraphics[clip=]{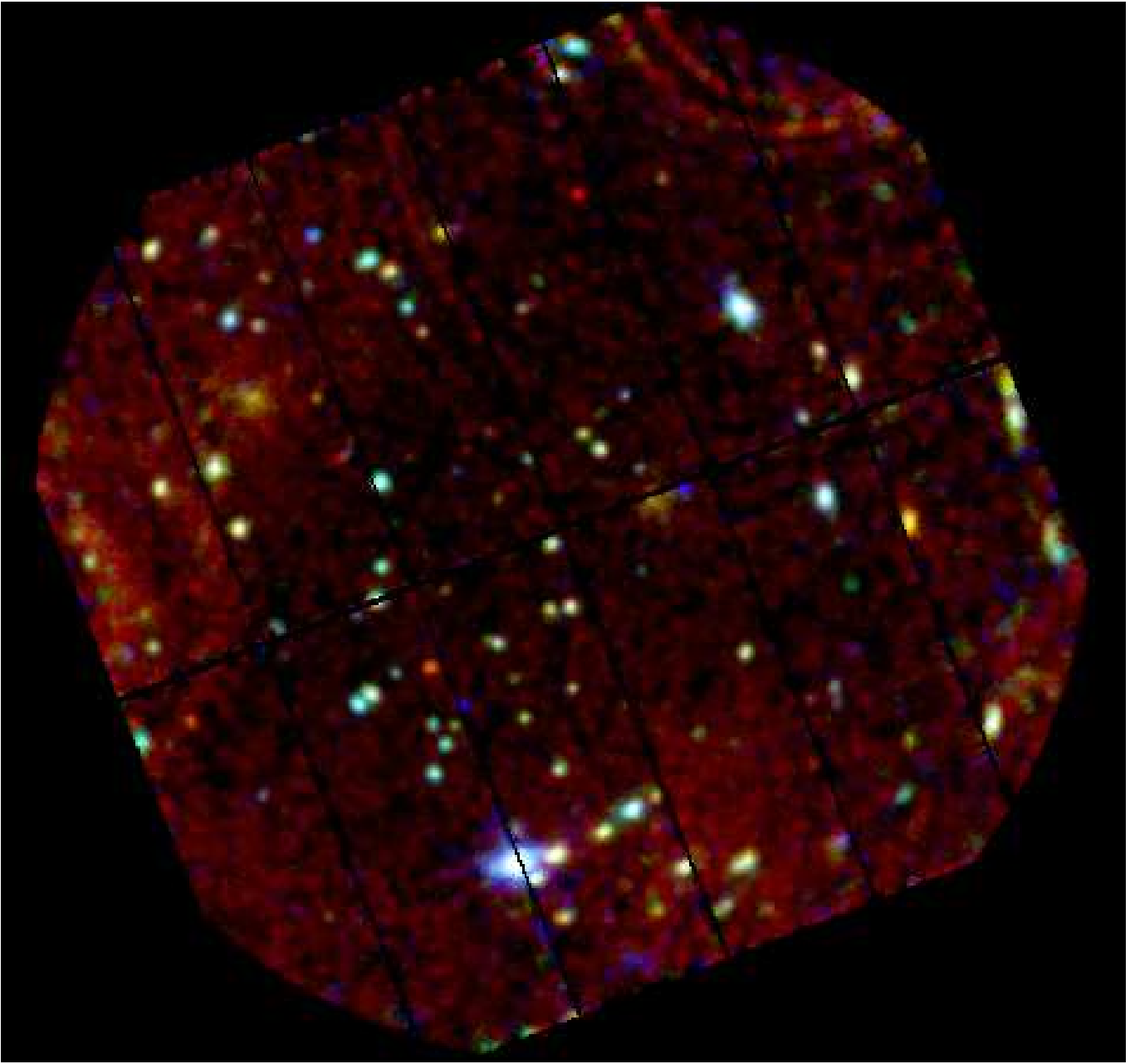}}
\caption{EPIC-pn RGB color image produced from data in three energy bands
         (red: 0.3--1.0 keV, green: 1.0--2.0 keV and blue: 2.0--7.5 keV).
         Each individual image was smoothed with an adaptive intensity 
         filter. Instrumental
         background from an observation with closed filter wheel (from
         revolution 59) was subtracted before vignetting and exposure 
         correction. The XMM-Newton pointing was aimed at 
	 RA = 05$^{\rm h}$31$^{\rm m}$20$^{\rm s}$
         and Dec = --65\degr57\arcmin38\arcsec\ and covers 
         about 13\arcmin\ in radius. North is to the top and West to 
         the right. The brightest source, located in the South, is the Be/X-ray
	 binary system \exo.}
\label{epic-ima}
\end{figure*}

\section{The XMM-Newton observation}

XMM-Newton \citep{2001A&A...365L...1J} observed the field located in the 
northern part of the LMC during satellite revolution 
\#152 on Oct. 7, 2000 from 03:36 to 22:43 UT with a gap in the middle
of about 90 min due to
ground station hand-over. A preliminary analysis of the EPIC data, which did not 
apply proper corrections for telescope vignetting and point spread function,
was reported in \citet{2002ESASP488H}.
Here results of the analysis of a selected sample 
of detected X-ray sources are presented utilizing the data collected with
the European Photon Imaging Cameras (EPICs). The cameras are
based on MOS \citep[EPIC-MOS1 and \hbox{-MOS2,}][]{2001A&A...365L..27T}
and pn \citep[EPIC-pn,][]{2001A&A...365L..18S} CCD detectors and
are mounted behind the three X-ray multi-mirror systems \citep{2000SPIE.4012..731A}
and observe simultaneously. All cameras were
operated in Full-Frame read-out mode providing data over a field of view of
$\sim$13\arcmin\ radius and used the medium filter to block out optical light. 
The data were processed using the XMM-Newton analysis package 
SAS version 5.3.3 to produce the photon event files and binned data products
like images, spectra and light curves. 

\subsection{The LMC field}

Images were created in the energy bands 0.3--1.0 keV (soft), 1.0--2.0 keV 
(medium) and 2.0--7.5 keV (hard) and smoothed using intensity-dependent 
adaptive filtering.
Detector-intrinsic particle induced background from an observation with 
closed filter wheel (on April 5, 2000) 
was subtracted before each image was corrected for telescope 
vignetting and exposure. A true colour image was produced by combining the 
three images and using red, green and blue
colour scales to represent the intensities in the soft, medium and hard
energy bands, respectively. This RGB image is presented in Fig.~\ref{epic-ima}.
According to their spectral energy distribution X-ray sources of different 
type are characterized by different colours. Sources predominantly radiating 
in the soft band appear in red while hard sources are blue.
Nearby bright sources outside the field of view (an order of magnitude brighter 
than the brightest source inside the field) cause arc-like structures 
in the images due to photons which are single-reflected on the mirror shells.
Arcs caused by the bright star AB\,Dor (at $\sim$35\arcmin\ angular distance 
from the optical axis) can be seen in the NW part of the
image. The supernova remnant N63A (an extended source $\sim$27\arcmin\ off-axis) 
located East produces 
extended arcs on the left side of the image. Another arc from the pair of 
bright SNRs (N49/N49B, $\sim$34\arcmin) is visible on the edge of the image to 
the SW. To first
order the energy dependence of the single-reflection efficiency is similar to
that of ``normal" double-reflected photons and the spectrum is little
altered. Therefore, the arcs from the SNRs show the typical red-green colours
of this class of sources with thermal spectra.

\begin{table*}
\caption[]{Sample of selected sources detected in the EPIC observations.}
\begin{tabular}{lcrrrrcrl}
\hline\noalign{\smallskip}
\multicolumn{1}{l}{Source} &
\multicolumn{1}{c}{ML} &
\multicolumn{1}{c}{PSPC} &
\multicolumn{1}{c}{HRI} &
\multicolumn{1}{c}{R} &
\multicolumn{1}{c}{B} &
\multicolumn{1}{c}{USNO} &
\multicolumn{1}{c}{delta} &
\multicolumn{1}{c}{Comment} \\
\multicolumn{1}{l}{XMMU\,J05...} &
\multicolumn{1}{c}{} &
\multicolumn{1}{c}{} &
\multicolumn{1}{c}{} &
\multicolumn{1}{c}{[mag]} &
\multicolumn{1}{c}{[mag]} &
\multicolumn{1}{c}{U0225\_02...} &
\multicolumn{1}{c}{[\arcsec]} &
\multicolumn{1}{c}{} \\

\noalign{\smallskip}\hline\noalign{\smallskip}
 2924.3-655724 & 1.60\expo{3} & 210 & 190 &      &      &        &     & foreground star? GSC8891.0505 13.68mag\\
 2940.5-655318 & 1.38\expo{3} & 189 & 198 & 17.9 & 19.0 & 041191 & 2.7 & AGN?\\
 2947.4-655639 & 2.26\expo{3} & 204 & 202 & 15.0 & 14.3 & 044134 & 4.3 & HMXB-Be RX J0529.8-6556 69.5 s pulsar\\
 3011.2-655122 & 1.61\expo{4} & 183 & 205 & 14.7 & 14.2 & 054120 & 1.5 & HMXB-Be? 272 s pulsar\\
 3034.9-655653 & 1.27\expo{2} & 205 &     &      &      &        &     & SNR?\\
\noalign{\smallskip}
 3041.0-660532 & 3.07\expo{3} &     & 210 & 17.3 & 18.1 & 066312 & 3.3 & AGN?\\
 3049.9-655522 & 1.12\expo{3} & 197 &     & 17.8 & 18.5 & 070177 & 2.8 & AGN?\\
 3056.2-654809 & 1.49\expo{2} &     &     & 17.9 & 19.4 & 072699 & 1.6 & SSS?\\
 3102.5-660649 & 3.52\expo{3} &     & 215 &      &      &        &     & AGN?\\
 3113.3-660705 & 4.04\expo{4} & 252 & 218 & 13.9 & 14.7 & 079445 & 2.5 & HMXB-Be \exo\ 13.7 s pulsar\\
\noalign{\smallskip}
 3137.2-660131 & 4.20\expo{2} & 225 &     & 14.0 & 16.0 & 088686 & 2.2 & foreground star?\\
 3150.9-655616 & 3.76\expo{3} & 202 &     &      &      &        &     & AGN?\\
 3153.8-660217 & 1.71\expo{3} & 229 & 222 & 17.1 & 18.0 & 095134 & 3.5 & AGN?\\
 3153.9-654959 & 1.56\expo{3} &     &     &      &      &        &     & AGN?\\
 3157.1-660233 & 1.25\expo{3} &     &     & 17.9 & 18.7 & 096316 & 1.7 & AGN?\\
\noalign{\smallskip}
 3226.2-655352 & 3.19\expo{2} & 190 &     &      &      &        &     & SNR?\\
 3230.2-655735 & 1.43\expo{3} & 211 &     &      &      &        &     & AGN?\\
 3232.4-655139 & 7.84\expo{2} & 184 & 233 & 13.9 & 12.4 & 109716 & 2.0 & HMXB-OB RX J0532.5-6551\\
 3236.7-655551 & 2.15\expo{3} &     & 236 & 17.3 & 18.1 & 111244 & 4.8 & AGN?\\
 3243.9-660258 & 9.85\expo{1} &     &     & 11.7 & 14.4 & 113921 & 4.0 & foreground star?\\
\noalign{\smallskip}
\hline
\end{tabular}
\label{tab-sources}

Notes: Maximum Likelihood (ML) corresponding to probability P = 1--exp(-ML) of existence; PSPC
and HRI catalogue entries from \citep{1999A&AS..139..277H} and \citep{2000A&AS..143..391S}; R and B
from nearest (within 5\arcsec) USNO A2.0 catalogue entry; delta denotes the angular distance between
optical (USNO A2.0) and X-ray position
\end{table*}

\section{Selected X-ray sources}

\begin{figure}
\resizebox{\hsize}{!}{\includegraphics[clip=]{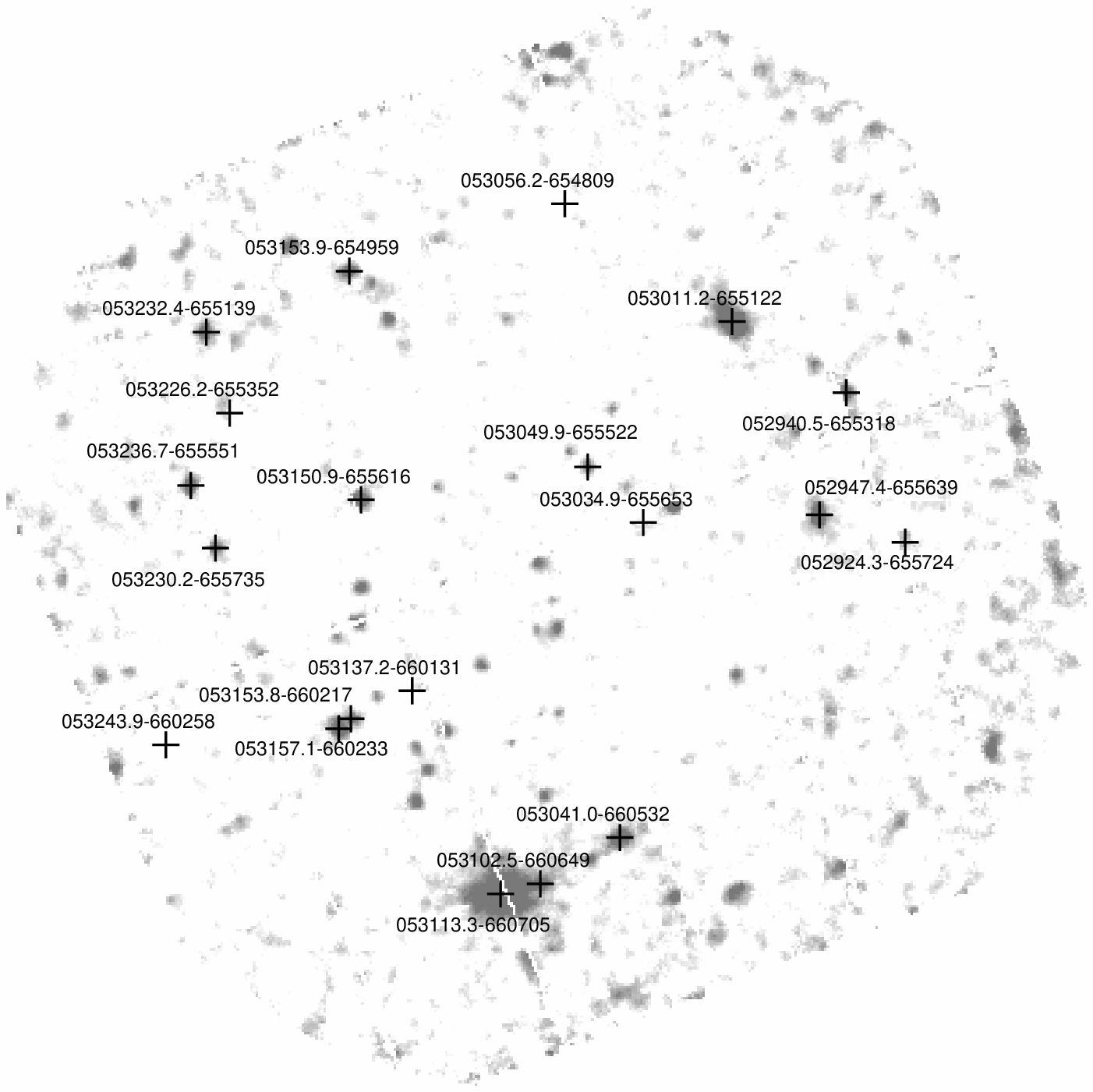}}
\caption{EPIC-pn 0.3--7.5 keV grey-scale image with the sources investigated in this 
         work marked.}
\label{ima-label}
\end{figure}

Source detection based on sliding window and maximum likelihood methods 
available in the SAS package yielded 
about 150 discrete X-ray sources in the field. A source list with X-ray 
properties and a statistical analysis will be presented in a follow-up paper. 
Here a sample of twenty sources is investigated which is selected on the basis of 
brightness, hardness and angular extent. To easily identify the sources in 
Fig.~\ref{epic-ima}, their locations are marked in a grey-scale broad-band image
in  Fig.~\ref{ima-label}. Some information useful for the identification of the 
sources is provided in Table~\ref{tab-sources}. Sixteen of them are located within 
15\arcsec\ distance to ROSAT sources as presented in 
\citet{1999A&AS..139..277H} and \citet{2000A&AS..143..391S} 
and thirteen have a likely optical counterpart in the 
USNO A2.0 catalogue (one additional bright star is listed only in the HST guide star 
catalogue). Angular distances between the X-ray and optical positions are listed in
the table (delta) together with the optical R and B magnitudes given in the USNO 
catalogue.

Finding charts produced from the DSS2 (red) image for some of the sources
presented here are published in \citet{2002ESASP488H}, however with preliminary 
X-ray source positions which can differ of the order of a few arc seconds.
Among the optically brightest objects in the sample three are likely 
late-type foreground stars given their red colours. The identification of the 
X-ray sources with these stars is supported by the X-ray spectrum (see below). 
Three other objects with B brighter than 15 mag are identified with known HMXBs 
and the fourth with \xmmbex, a candidate HMXB proposed by \citet{2002ESASP488H},
based on the X-ray spectrum and the brightness of the optical counterpart which is
similar to those of the known HMXBs in the field. \citet{2002ESASP488H} also proposed
two new candidates for SNRs which can be easily identified in Fig.~\ref{epic-ima}
from their angular extent and colours. For the brightest fourteen sources a 
sufficient number of counts ($>$500 in pn) were collected to allow spectral and 
temporal analyses. For sources with extreme X-ray spectrum a smaller number of counts 
allows a classification of the X-ray source. Six such sources were selected including 
a likely faint SSS, two SNRs, two more foreground stars and the known HMXB \rxob. 

\subsection{EPIC X-ray spectra}

Source spectra averaged over the whole observation were extracted from MOS and pn 
data in circular regions around the source 
positions with typical radii of 16\arcsec--22\arcsec. For the extended candidate 
SNRs larger radii of 30\arcsec\ and 45\arcsec\ were used. Background was taken 
from nearby circular regions chosen to be at a comparable distance to the 
read-out node of the pn detector (RAWY coordinate) in order to avoid
detector-intrinsic background variations. 
Redistribution matrices available with SAS\,5.3 were used in 
combination with effective area files produced by ``arfgen" which include 
corrections for vignetting and point spread function losses. The set of three 
spectra (MOS1, MOS2 and pn) was analyzed using XSPEC by simultaneously fitting 
a common spectral model where the normalizations between the instruments were 
independent parameters to allow for inter-calibration uncertainties. 
Errors and upper limits were derived for 90\% confidence levels. Except for 
the brightest source (\exo) which requires a more complex model, single 
component models attenuated by photo-electric absorption were fit to the 
spectra. EPIC spectra from eight representative sources from the 
sample are shown in Fig.~\ref{epic-spectra}. Various source types can be clearly 
distinguished from their different energy distribution in the 0.15--10.0 keV energy 
band.

\begin{table*}[t]
\caption[]{Spectral fit results.}
\begin{tabular}{lllllll}
\hline\noalign{\smallskip}
\multicolumn{1}{l}{Source} &
\multicolumn{1}{c}{$\gamma$} &
\multicolumn{1}{c}{kT} &
\multicolumn{1}{c}{\nh} &
\multicolumn{1}{c}{SM$^4$} &
\multicolumn{1}{c}{Flux$^5$} &
\multicolumn{1}{c}{Luminosity$^6$} \\

\multicolumn{1}{l}{XMMU\,J05...} &
\multicolumn{1}{c}{} &
\multicolumn{1}{c}{[keV]} &
\multicolumn{1}{c}{[\oexpo{21}cm$^{-2}$]} &
\multicolumn{1}{c}{} &
\multicolumn{1}{c}{[erg cm$^{-2}$ s$^{-1}$}] &
\multicolumn{1}{c}{[erg s$^{-1}$]} \\

\noalign{\smallskip}\hline\noalign{\smallskip}
 2924.3-655724 & & 0.68$\pm$0.08          & $<$0.1              & MK & 6.6\expo{-14} &\\
 2940.5-655318 & 2.33$^{+0.13}_{-0.20}$ & & 1.3$^{+0.2}_{-0.4}$ & PL & 9.6\expo{-14} &\\
 2947.4-655639 & 1.31$\pm$0.17          & & 2.6$\pm$1.0         & PL & 2.2\expo{-13} & 7.7\expo{34}\\
 3011.2-655122 & 1.07$\pm$0.06          & & 6.9$\pm$0.3         & PL & 1.2\expo{-12} & 3.9\expo{35}\\
 3034.9-655653 & & 0.7 fixed              & 1.9$^{+0.8}_{-0.6}$ & MK & 1.9\expo{-14} & 1.2\expo{34}\\
\noalign{\smallskip}
 3041.0-660532 & 1.40$^{+0.14}_{-0.11}$ & & 2.7$^{+0.9}_{-0.7}$ & PL & 2.9\expo{-13} &\\
 3049.9-655522 & 2.00$^{+0.20}_{-0.21}$ & & 1.4$^{+0.5}_{-0.5}$ & PL & 5.1\expo{-14} &\\
 3056.2-654809$^1$ & & 0.049$^{+0.006}_{-0.012}$ & $<$0.3           & BB & 4.2\expo{-14} & 1.3\expo{34}\\
 3102.5-660649$^2$ & 1.81$^{+0.09}_{-0.07}$ & & $<$0.1          & PL & 2.3\expo{-13} &\\
 3113.3-660705 & 1.67$^{+0.12}_{-0.13}$ & 0.10$\pm$0.02 & 6.9$^{+0.7}_{-1.3}$ / 97$^{+3}_{-7}$ & MC & 7.0\expo{-12} & 4.6\expo{37}\\
\noalign{\smallskip}
 3137.2-660131 & & 0.66$\pm$0.16          & $<$1.0              & MK & 1.6\expo{-14} &\\
 3150.9-655616 & 1.92$^{+0.14}_{-0.07}$ & & 5.3$^{+0.9}_{-0.7}$ & PL & 1.6\expo{-13} &\\
 3153.8-660217 & 2.07$^{+0.20}_{-0.17}$ & & 2.8$\pm$0.6         & PL & 1.0\expo{-13} &\\
 3153.9-654959 & 1.88$^{+0.30}_{-0.26}$ & & 3.8$^{+1.8}_{-1.2}$ & PL & 1.1\expo{-13} &\\
 3157.1-660233 & 1.62$^{+0.14}_{-0.17}$ & & 3.8$^{+1.0}_{-1.2}$ & PL & 1.4\expo{-13} &\\
\noalign{\smallskip}
 3226.2-655352 & & 0.68$\pm$0.06          & 2.1$^{+0.9}_{-0.6}$ & MK & 5.5\expo{-14} & 3.6\expo{34}\\
 3230.2-655735$^3$ & 2.23$^{+0.20}_{-0.13}$ & & 0.7$^{+0.3}_{-0.5}$ & PL & 6.3\expo{-14} &\\
 3232.4-655139 & 1.18$^{+0.25}_{-0.29}$ & & 2.5$^{+2.2}_{-1.5}$ & PL & 1.2\expo{-13} & 4.1\expo{34}\\
 3236.7-655551 & 2.02$^{+0.19}_{-0.16}$ & & 1.3$^{+0.5}_{-0.4}$ & PL & 1.1\expo{-13} &\\
 3243.9-660258 & & 0.52$\pm$0.18          & $<$0.4              & MK & 1.0\expo{-14} &\\
\noalign{\smallskip}
\hline
\end{tabular}
\label{tab-fits}

$^1$ spectral fit to pn spectrum only\\
$^2$ the low absorption indicates a low energy excess which is probably caused by contamination from \exo\\
$^3$ located in arc structure caused by single reflected photons from nearby bright SNR which 
     probably causes contamination\\
$^4$ Spectral model; MC: multi-component (see text), PL: power-law, BB: blackbody, MK: thermal emission from plasma in
     collisional equilibrium \citep[Mewe-Kaastra,][]{1985A&AS...62..197M}\\
$^5$ 0.2--10 keV, determined from the pn spectrum except for \exo\ (MOS1) which is
     located on a CCD gap in the pn\\
$^6$ 0.2--10 keV, intrinsic luminosity with \nh\ set to 0 for sources located in the LMC assuming 
     a distance of 50 kpc
\end{table*}

\subsubsection{High Mass X-ray Binaries}

Although the Be/X-ray pulsar \exo\ is known since almost 20 years, little is 
known about its X-ray spectrum. The small angular separation 
to LMC\,X-4 made it impossible to take spectra with non-imaging instruments
and only measurements with the EXOSAT Low Energy imaging detector with 
different filters at energies below 2 keV were available \citep{1987PhD.Brunner}.
In the 2--20 keV band a spectrum of low statistical quality was obtained from the 
coded mask SL2 XRT experiment which was consistent with a power-law with a photon 
index 1.0$^{+0.6}_{-0.5}$ and an absorbing hydrogen column density of 
7$^{+16}_{-7}$\hcm{22} \citep{1989MNRAS.240P...1H}. 

\begin{figure*}
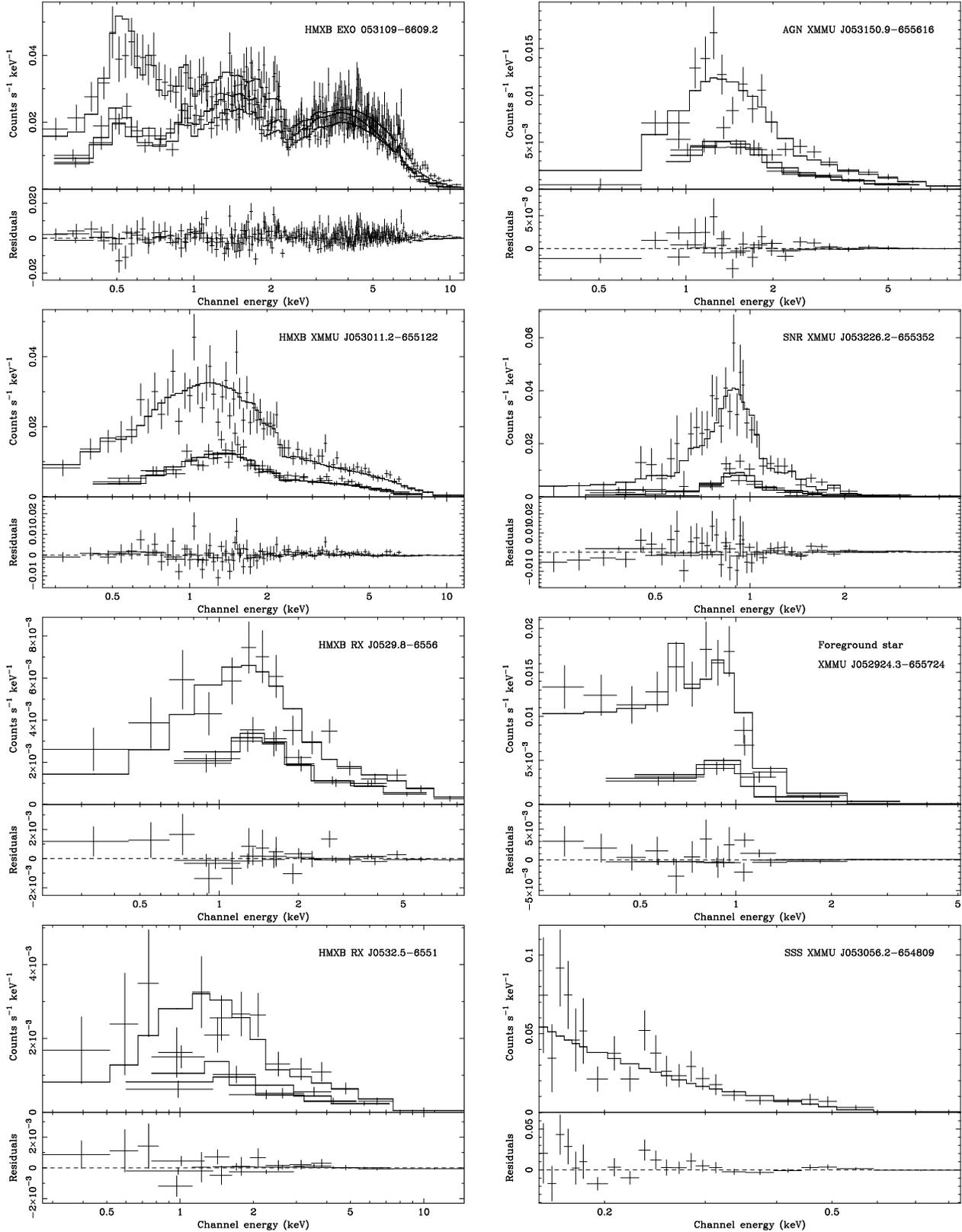

\hbox{
\resizebox{8cm}{!}{\includegraphics[clip=,angle=-90]{p010406pns001_sc1_pow.ps}}
\hspace{5mm}
\resizebox{8cm}{!}{\includegraphics[clip=,angle=-90]{p010406pns001_sc4_pow.ps}}
}
\hbox{
\resizebox{8cm}{!}{\includegraphics[clip=,angle=-90]{p010406pns001_sc2_pow.ps}}
\hspace{5mm}
\resizebox{8cm}{!}{\includegraphics[clip=,angle=-90]{p010406pns001_sc38_mekal.ps}}
}
\hbox{
\resizebox{8cm}{!}{\includegraphics[clip=,angle=-90]{p010406pns001_sc8_pow.ps}}
\hspace{5mm}
\resizebox{8cm}{!}{\includegraphics[clip=,angle=-90]{p010406pns001_sc11_mekal.ps}}
}
\hbox{
\resizebox{8cm}{!}{\includegraphics[clip=,angle=-90]{p010406pns001_sc21_pow.ps}}
\hspace{5mm}
\resizebox{8cm}{!}{\includegraphics[clip=,angle=-90]{p010406pns001_sc64pn_bb.ps}}
}
\caption{EPIC spectra of selected X-ray sources in the LMC field. For each source 
the best fit to the three EPIC spectra is shown in the upper panels (pn is 
always the upper spectrum and the two MOS spectra overlap). The four HMXBs 
including the new candidate \xmmbex\ are shown on the left.
The right column illustrates the variety of X-ray spectra from sources of 
different nature. The spectra of candidate AGN show steeper power-laws
than HMXBs. Spectra of the candidate SNR \xmmsnr\ and foreground stars 
exhibit little flux above 2 keV but can be distinguished by the amount of 
absorption. The softest spectrum is detected from the new candidate SSS 
\xmmsss\ which can be represented by a blackbody model with kT$\sim$50 eV. 
For this case only the pn data yields sufficient statistics to accumulate 
a spectrum.
}
\label{epic-spectra}
\end{figure*}

The XMM-Newton observation yielded the first 0.2--12 keV spectrum 
(Fig.~\ref{epic-spectra}) of \exo. The spectrum is complex and cannot be 
represented by the simple power-law (neither by including a high-energy cutoff) 
model sufficient to characterize most Be/X-ray binaries. Below 1 keV clearly 
a soft component is 
detected with features too narrow to be fit with continuum spectra as produced 
by e.g. blackbody or multi-temperature disk-blackbody emission. The only 
acceptable model was found to be thermal emission from hot plasma in collisional 
equilibrium \citep[VMEKAL model in XSPEC,][]{1985A&AS...62..197M}
with a temperature kT of 0.1 keV and 
reduced abundances (0.29 solar) of elements heavier than oxygen. Above 3 keV 
a highly absorbed ($\sim$\ohcm{23}) power-law gives an acceptable fit consistent 
with the SL2 results. 
However, both components described so far do not contribute significantly to
the flux between 1 and 2 keV. Reducing the absorption in the power-law component
yields too high flux around 2--3 keV and does not fit. Several components were
tested in the model to account for the emission between 1 and 2 keV. Neither
blackbody, disk-blackbody nor a second VMEKAL give acceptable results. 

\begin{figure}
\resizebox{8cm}{!}{\includegraphics[clip=,angle=-90]{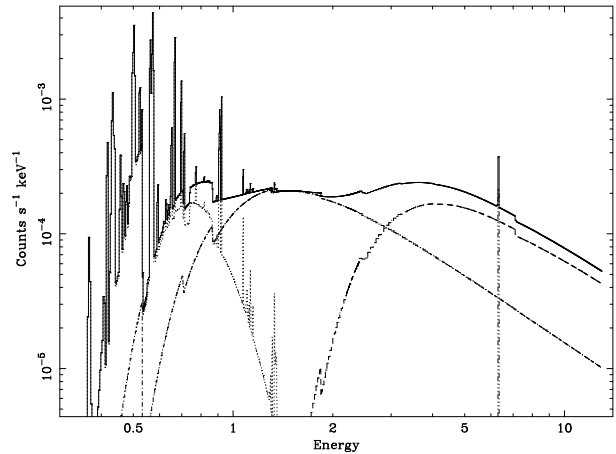}}
\caption{Best fit model spectrum for \exo. It comprises of two power-law components 
with equal slope absorbed by different amounts of column density (``partial absorber") 
and a thermal plasma emission component. For parameters see Table~\ref{tab-fits}.}
\label{exo-model}
\end{figure}

The best 
fit was obtained by adding another power-law component which is attenuated at low 
energies by the same amount of absorption as the thermal plasma component.
The power-law index was forced to be the same as the index of the high absorption
power-law. The best fit values for the photon index and the two column densities 
are listed in Table~\ref{tab-fits}. The normalization of the low-absorption component
relative to the high-absorption component is 0.23.
The two power-law components may originate from the
same emission region partially covered by different amounts of absorbing matter.
Partially may be interpreted here as spatially or temporally varying. The latter is
supported by spectral variations seen during the pulse period (see below).
Adding a narrow Fe emission line (6.32$\pm0.06$ keV) with an equivalent width of 
329 eV improved the $\chi^2$ by 13 to the best fit with $\chi^2$=520 for 336 degrees 
of freedom. The model spectrum is illustrated in Fig.~\ref{exo-model}.

A soft spectral component is so far only detected from a few HMXBs, most of them
located in the Magellanic Clouds
\citep[e.g.][]{2000PASJ...52..299K,2000ApJ...539..191Y}. 
In the Milky Way the HMXBs are concentrated in the galactic plane and the large 
interstellar absorption strongly attenuates the soft X-ray energies.
Therefore, the HMXBs in the Magellanic Clouds are ideal to investigate the soft part
of the spectrum. No pulsations were detected in the soft spectral component seen from
RX\,J0059.2-7138 with ASCA \citep{2000PASJ...52..299K} which suggests emission from a
relatively large region. The emission measures derived for \exo, if the thin plasma
model is adopted, of $\sim4$\expo{61} cm$^{-3}$ and for RX\,J0059.2-7138 of 
$\sim$\oexpo{61}cm$^{-3}$
\citep{2000PASJ...52..299K} support this. A possible origin of the soft emission may
be the disk-like dense stellar wind around the Be star which is illuminated by the 
strong X-ray source with luminosities of the order of \oergs{38}. However, to 
differentiate between emission lines produced in collisionally or photo-ionized
plasma high resolution grating spectra are required. E.g. Chandra observations
of the HMXB Cen\,X-3, a close system with supergiant primary, revealed emission 
lines excited by both kinds of processes \citep{2002astro.ph..6065W}. 

From \rxob\ and \rxbe\ only ROSAT PSPC spectra in the energy band 0.1--2.4 keV were 
available. The spectra are of low statistical quality and the narrow energy coverage
did not allow to determine power-law index and column density with high accuracy
\citep{1995A&A...303L..49H,1997A&A...318..490H}.
\citet{2002ESASP488H} proposed a new candidate HMXB based on its hard X-ray 
spectrum and the presence of a likely optical counterpart on the DSS2 (R) 
image with similar brightness as the counterparts of the other two nearby 
Be/X-ray binaries. Detection of a possible pulse period 
(see below) for \xmmbex\ leaves little doubt about the HMXB and most 
likely Be/X-ray binary nature of this source. 
Fitting an absorbed power-law to the spectra of \xmmbex, \rxbe\ and \rxob\ yields
acceptable fits (reduced $\chi^2$ 0.9--1.3) with power-law indices in the range
1.0--1.4 (Table~\ref{tab-fits}). The EPIC spectra of these sources are 
shown together with the best fit model in Fig.~\ref{epic-spectra}. 
\xmmbex\ was also detected in ROSAT data 
\citep{1999A&AS..139..277H,2000A&AS..143..391S} with a count rate corresponding to a
flux which is a factor of 2.2 below the XMM-Newton measurement (assuming spectral
parameters derived from the EPIC spectra).

\subsubsection{AGN}

The majority of the X-ray brightest sources in the field (ten out of fourteen) 
most likely consists 
of AGN in the background of the LMC. Their X-ray spectra can be represented by 
power-laws with indices between 1.8 and 2.3 with only one case with index 1.4 
(Table~\ref{tab-fits}), typically for this source class. Moreover they do
not possess an optical counterpart with B brighter than 18th magnitude in the USNO
A2.0 catalogue (Table~\ref{tab-sources}). As an example the EPIC spectra of 
XMMU\,J053150.9-655616 together with the best fit model are plotted in 
Fig.~\ref{epic-spectra}.

\subsubsection{Supernova remnants}

Two new SNRs were proposed by \citet{2002ESASP488H} from their spatial extent 
and X-ray colours (Fig.~\ref{epic-ima}) in the EPIC data. 
Both SNRs were detected in the ROSAT PSPC data 
\citep[sources 190 and 205 in][]{1999A&AS..139..277H}
with indication for spatial extent. However, due to the low statistical quality of 
the data the sources were not classified as SNR candidates.
\xmmsnr\ is sufficiently bright to
allow the investigation of the X-ray spectrum which shows little flux above 2
keV. Fitting a MEKAL model results in a best fit temperature kT = 0.68 keV
consistent with radiation from a thermal SNR. For the Fe abundance a best fit
value of 0.41$^{+0.15}_{-0.11}$ was derived. The best fit model to the EPIC 
spectra of \xmmsnr\ is shown in Fig.~\ref{epic-spectra}. Due to the lower number 
of detected photons from  XMMU\,J053034.9-655653 the temperature was fixed at 
0.7 keV in the fit to the EPIC spectra. Also in this case a low Fe abundance 
was deduced (upper limit of 0.06). 

\subsubsection{Foreground stars}

The EPIC spectra of sources identified with foreground stars also do not 
show significant emission above 2 keV. However the absorbing column density is
lower than that of sources located in the LMC clearly distinguishing them from SNRs
(in addition they are unresolved X-ray sources). Spectra of three foreground stars
were investigated and the results are listed in Table~\ref{tab-fits}. All three 
spectra can be represented by emission from hot thermal emission with temperature 
between 0.5 and 0.7 keV and low Fe abundance of 0.1--0.2 solar. For the absorption 
column densities only upper limits could be derived consistent with nearby objects.
The X-ray spectra suggest emission from late-type stars with active corona which is also
consistent with their B-R index derived from the magnitudes listed in USNO A2.0.
As an example the spectra of XMMU\,J052924.3-655724 are plotted in 
Fig.~\ref{epic-spectra}.

\subsubsection{Supersoft source}

The faint source \xmmsss\ is characterized by red colour in the image of 
Fig.~\ref{epic-ima} indicating an extremely soft X-ray spectrum. Only the pn 
detector is sensitive enough to allow accumulation of an energy spectrum which 
is only detected below 0.5 keV (see Fig.~\ref{epic-spectra}). 
A formal fit with a blackbody model yields a temperature
of 49 eV which is typical for supersoft sources, AM Her systems, isolated white
dwarfs or isolated X-ray dim neutron stars. Its luminosity of 1.3\ergs{34}
for a distance of 50 kpc is too high to originate from an AM Her type
cataclysmic variable in the LMC \citep[for luminosities of the soft blackbody
component in AM Her systems see e.g.][]{1994MNRAS.270..692R}. 
\xmmsss\ is not contained in the ROSAT catalogues of the LMC and therefore archival 
PSPC data (45 ks observation with ID 200692P) was investigated and an 
upper limit of 7.9\expo{-4} \ct\ (3$\sigma$, 0.1--0.6 keV) was determined.
Folding the spectral model fitting the EPIC pn data through the PSPC response an
expected count rate of 2.6\expo{-2} \ct\ is obtained which implies that 
\xmmsss\ was at least a factor of 33
fainter during the ROSAT observation in October/November 1991. The high 
variability excludes an explanation as dim isolated neutron star or single 
white dwarf in the foreground. An AM Her system located in the Milky Way 
is expected to be brighter in the optical than the close USNO catalogue object. 
Therefore, \xmmsss\ is considered most likely to be a variable SSS in the LMC.

\subsection{X-ray pulsars}

Temporal analysis was performed on the data of the HMXBs and the other 
eleven brightest sources in the sample. Fast Fourier power spectra were produced from
light curves in the 0.3--7.5 keV energy band. A folding analysis was then
performed around frequency peaks identified in the power spectra. Pulsations were 
detected only from the known HMXB pulsars and from the new candidate HMXB which is 
discussed in the following.

\subsubsection{\exo}

\begin{figure*}
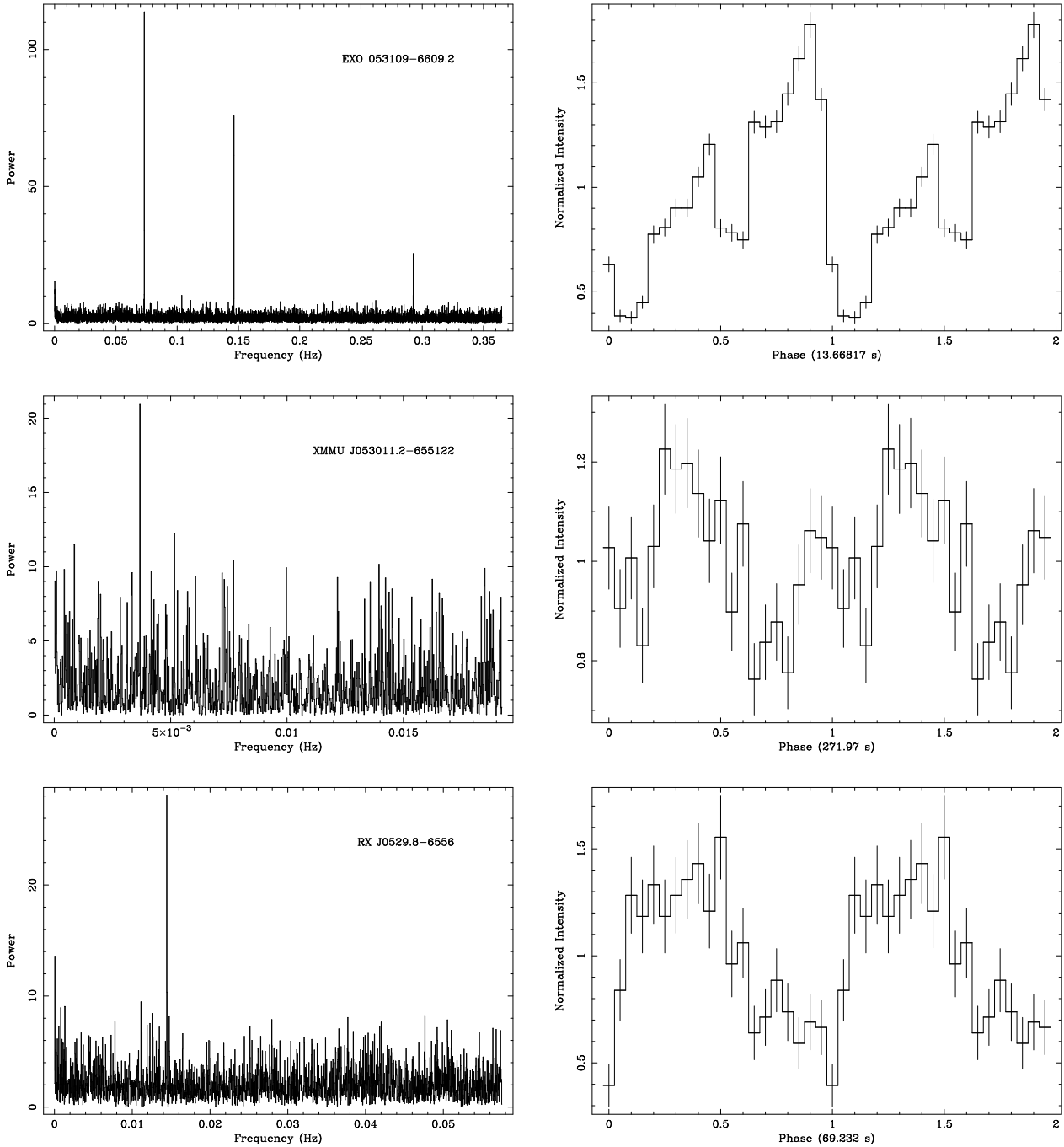

\hbox{
\resizebox{8cm}{!}{\includegraphics[clip=,angle=-90]{powspec_sc1_300_7500.ps}}
\hspace{5mm}
\resizebox{8cm}{!}{\includegraphics[clip=,angle=-90]{efold_sc1_300_7500.ps}}
}
\vspace{5mm}
\hbox{
\resizebox{8cm}{!}{\includegraphics[clip=,angle=-90]{powspec_sc2_300_7500.ps}}
\hspace{5mm}
\resizebox{8cm}{!}{\includegraphics[clip=,angle=-90]{efold_sc2_300_7500.ps}}
}
\vspace{5mm}
\hbox{
\resizebox{8cm}{!}{\includegraphics[clip=,angle=-90]{powspec_sc8_300_7500.ps}}
\hspace{5mm}
\resizebox{8cm}{!}{\includegraphics[clip=,angle=-90]{efold_sc8_300_7500.ps}}
}
\caption{Power spectra and pulse profiles of HMXB pulsars produced from EPIC-pn 
data in the 0.3--7.5 keV energy band.}
\label{hmxb-pulse}
\end{figure*}

For \exo\ a mean pulse period of 13.66817(1) s was derived (1$\sigma$ error given for 
the last digit). The power spectrum
exhibits also peaks at the first and third harmonic, consistent with the highly
non-sinusoidal pulse profile (Fig.~\ref{hmxb-pulse}). Pulsations are seen above
and below 4 keV confirming that the two spectral power-law components 
(Fig.~\ref{exo-model}) originate
both from \exo. Due to the strong re-distribution of events to lower 
energy-channels at energies below $\sim$1 keV in the EPIC pn CCD 
it is not possible to verify if the
soft thermal plasma emission below 0.5 keV shows pulsations or not. For a spectrum 
steeply falling towards lower energies the contribution from photons above 0.5 keV
dominates the events below channel 100 (nominally corresponding to energy 0.5 keV)
and therefore pulsations seen in the flux above 0.5 keV will also be seen in the
lowest channels.

\begin{figure*}
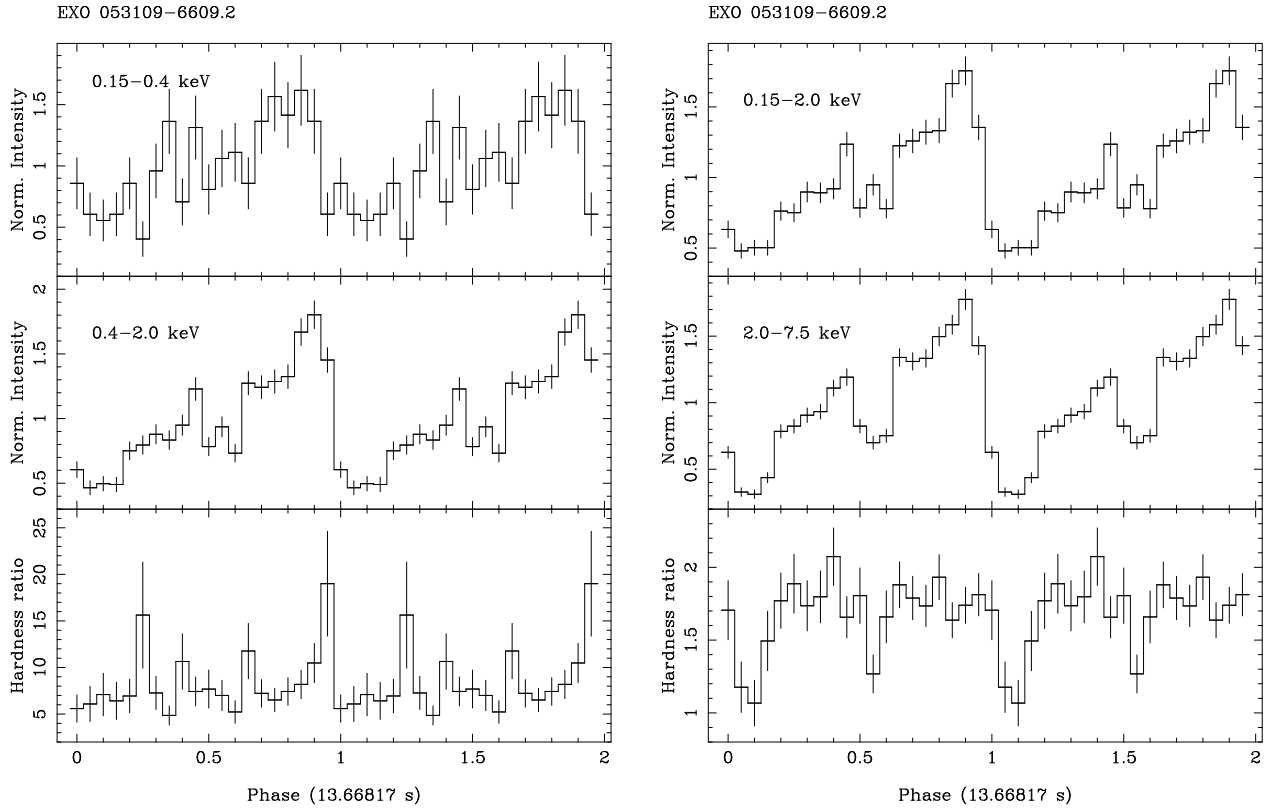

\hbox{
\resizebox{8cm}{!}{\includegraphics[clip=]{efold_sc1_150_400_2000.ps}}
\hspace{5mm}
\resizebox{8cm}{!}{\includegraphics[clip=]{efold_sc1_150_2000_7500.ps}}
}
\caption{Folded light curves obtained from EPIC pn data from \exo\ in 
different energy bands together with the hardness ratios (count rates 
in hard band / count rates in soft band).}
\label{exo-pulse}
\end{figure*}

To investigate possible spectral changes during the pulse period,
hardness ratio plots were produced by creating light curves folded on the pulse
period in different
energy bands and dividing the obtained count rates as function of pulse phase.
The folded light curves from energy bands 0.15--0.4 keV and 0.4--2.0 keV are 
plotted in Fig.~\ref{exo-pulse}~(left) together with the hardness ratio. The 
hardness ratio shows remarkable increases around the two intensity dips which
resemble the behaviour of an X-ray source viewed through a dense atmosphere before
and after it is eclipsed. The facts that the eclipse seen here is not total and 
that two eclipses are seen about 0.5 in phase apart, 
suggest that we see two emission regions in \exo, most likely from
the accreting magnetic poles of the neutron star. Before the pole is self-eclipsed 
by the neutron star or the accretion column the emission is viewed through dense 
matter, most likely matter
from the accretion column above the pole which attenuates the emission more 
strongly at lower
energies. During eclipse (when the emission is probably dominated by the other pole) 
the hardness ratio remains constant within the errors. 
Hardness ratios which involve emission from energies above 2 keV (which is less affected by
absorption and more sensitive to changes in the intrinsic spectrum) are plotted in
Fig.~\ref{exo-pulse}~(right). They show that the spectrum softens during the 
eclipses. This can be explained by a spatial temperature variation in the 
emission region with the temperature increasing with height above the neutron star
surface. The cooler regions become only visible during closest approach of the 
magnetic axis to the line of sight.
Such a picture is expected from matter in the accretion column shocked at a certain
height above the neutron star surface and cooling during the final settlement onto
the neutron star.

Comparison of the pulse profiles of the XMM-Newton observation with those obtained
from BeppoSAX data \citep{1998ApJ...498..831B} shows a major difference which is best
visible in the 2--10 keV band. During the BeppoSAX observation the profile shows a strong
peak after the deeper eclipse while after the secondary eclipse the intensity remains as
low as during the eclipse. During the XMM-Newton observation this behaviour is mirrored
around the secondary eclipse (at phase $\sim$0.55 in Fig.~\ref{exo-pulse})
with low intensity before and high intensity after the
eclipse. This behaviour is probably related to the overall source intensity.
The pulse profile obtained during a long ROSAT observation 
\citep{1996rftu.proc..131D} when the source was in a low-intensity state (1\ergs{35}) 
resembles those of the XMM-Newton observation (with observed luminosity of 1.7\ergs{35} 
in the 0.1--2.4 keV band). During the BeppoSAX observation the source was in outburst 
with an observed luminosity in excess of \oergs{37} compared to 2.1\ergs{36} measured 
by EPIC (note the luminosity given in Table~\ref{tab-fits} is corrected for absorption).
It should be noted that no attempt was made in the presented analysis to determine 
secular period changes and the pulse profiles based on a mean pulse period may be smeared out
somewhat.

\subsubsection{\rxbe\ and \xmmbex}

The pulse period of 69.232(2) s detected in the EPIC data from \rxbe\ confirms the
discovery by ROSAT. For the second brightest X-ray source in the field during the XMM-Newton
observations, \xmmbex, probable pulsations were found with a period of 
271.97(5) s. From the folding analysis the $\chi^2$ test yields a formal 
significance of 4.6 $\sigma$ for the probability that this detection is real.
Together with the X-ray spectrum and the brightness of the proposed optical 
counterpart the pulsations strongly support the identification of \xmmbex\ as fourth
HMXB system in the observed field.
Power spectra and light curves folded with the pulse period for \rxbe\ and \xmmbex\
are shown in Fig.~\ref{hmxb-pulse}.

\section{Source populations}

\citet{1997AJ....113.1815C} used a sample of SNRs in OB associations in the LMC to study
how interstellar environments affect the physical properties of SNRs. She found that SNRs
in \HII\ regions all show the three classical SNR signatures (bright X-ray emission, 
non-thermal radio emission and enhanced [\SII]/\Halp\ ratio) while some SNRs in
super-bubbles show only X-ray emission. The observed LMC field presented here is located 
at the northern rim of the SGS LMC\,4  
and one reason that the two new X-ray selected SNRs
were not detected yet at radio or optical wavelengths may be the explosion of their 
progenitors into the hot and low density interior of the SGS. Very likely the
formation of the progenitors was triggered by the expanding shell and the SNRs are now 
shocking the inner walls of the shell \citep{1997AJ....113.1815C}. A relatively 
large number of SNRs with X-ray luminosities around \oergcm{34} may therefore still be 
detectable in XMM-Newton observations of the Magellanic Clouds and in particular inside 
the supergiant shells in the LMC. Promising candidates may be found among the sources 
listed in the ROSAT catalogues 
\citep{1999A&AS..139..277H,2000A&AS..142...41H,2000A&AS..143..391S,2000A&AS..147...75S}
with indication for spatial extent but without sufficient number of detected photons 
(and therefore a low likelihood for the extent) to allow a classification of the sources.

The SMC is peculiar in its high number of Be/X-ray binaries 
\citep[both in absolute terms given the size of the galaxy and relative to the number 
of supergiant HMXBs and SNRs,][]{2000A&A...359..573H,2002Yokogawa} 
which may indicate an increased star forming rate about \oexpo{7} years ago
\citep[the time which elapses between the formation of a massive binary and its evolution 
into a HMXB,][]{1983vdHeuvel}. The star formation must have largely declined since then as the 
lower number of descendants of young massive stars like supergiant HMXBs and type-II SNRs 
indicates \citep{2002Yokogawa}. With respect to the space density of HMXBs the observed 
XMM-Newton field and its close neighborhood in the northern area of the SGS 
LMC\,4 are similar to that of the SMC. However, the existence of a relatively high number of 
younger SNRs and the supergiant 
HMXB system \rxob\ is consistent with a more constant star formation rate over the 
last \oexpo{7} years in this part of the LMC\,4 region. The stochastic self-propagating 
star formation (SSPSF) scenario proposed to explain the production of supergiant shells 
\citep{1981A&A....98..371F} 
predicts a gradient in the age of stars, from young stars at the edges to older 
ones inside. However, no such gradient was found for stars in the interior of LMC\,4 with 
stellar ages in a narrow range of 9 Myr to 16 Myr \citep{1997A&A...328..167B} 
leading to the conclusion that LMC\,4 may have formed without contribution from 
SSPSF, although the ring of young associations and \HII\ regions around the edge 
were triggered by the events inside the SGS. The X-ray source census of 
HMXBs and SNRs in the LMC\,4 region is fully consistent with this picture
and the Be/X-ray binaries may well belong to the stellar population found by 
\citet{1997A&A...328..167B}.

The majority of X-ray sources in the observed field are most likely AGN behind the LMC.
Their spectra are characterized by power-laws attenuated by photo-electric absorption
due to interstellar gas in the Milky Way and the LMC along the line of sight. 
From \Hone\ 21-cm maps the Galactic 
contribution is measured to 5.7\hcm{20} \citep{1990ARAA...28..215D} while the LMC column 
density varies between 5\hcm{20} and 9\hcm{20} \citep{1998ApJ...503..674K} across the field.
From the ten AGN candidates about half have X-ray measured column densities well above the
\Hone\ values. Since the absorption in the X-ray band is dominated by the metals in the 
absorbing gas lower metallicities adopted in the spectral fits 
\citep[0.5 solar for the LMC,][]{1992ApJ...384..508R} 
would lead to correspondingly higher hydrogen column densities.
Similar results, but with larger excess column densities, were obtained for AGN detected 
in the 30 Doradus region and 
\citet{2001A&A...365L.208H} concluded for the 30 Doradus region that either large amounts 
of \HII\ and/or \Hmol\ or higher metal abundances exist along the line of sights 
through the LMC. The case of the XMM-Newton field presented here is less clear.
No dense clouds are visible on \Halp\ images \citep[e.g.][]{1995A&A...296..523D} or
CO maps \citep{2001PASJ...53..959Y} in the observed field at the northern rim
of the LMC\,4 SGS and in the sample of candidate AGN the excess absorption
may be explained by AGN intrinsic obscuration. Indeed relatively large 
fractions of obscured AGN were found in deep surveys \citep[e.g.][]{2002astro.ph..7166M},
however, mainly among the X-ray and optically fainter ones. The ten brightest AGN in the 
LMC field have 2.0--10.0 keV fluxes higher than 3\ergcm{-14} and six of them likely
counterparts with R magnitudes (from USNO A2.0) between 17 and 18, confining them to 
the upper left part of the flux -- magnitude diagram of Fig.\,6 in
\citet{2002astro.ph..7166M}. There objects are located which show a much lower 
fraction of obscured AGN. To help to identify those unobscured AGN which can be used to probe 
the interstellar medium in the LMC optical spectra are required.

\section{Summary}

The deep XMM-Newton observation of a northern field in the LMC allows to uniquely 
classify the detected X-ray sources to flux levels of $\sim$\oergcm{-14} from their 
X-ray properties alone. X-ray spectra obtained by the EPIC cameras over a relatively 
broad energy band (0.15--12.0 keV) emphasize the various energy distributions observed 
from the different types of X-ray sources. A first analysis of twenty selected sources,
presented here, yielded the discovery of a previously undetected supersoft source, 
revealed two ROSAT sources with indication for spatial extent as supernova remnants 
and rendered the detection of a possible pulsation in the X-ray flux of a 
candidate HMXB in the LMC. 
This increases the number of known HMXBs in the observed field to four. 
Among the fourteen brightest sources three HMXBs, one foreground star and up to ten 
AGN were found. Particular results on individual objects are discussed in 
Sect.\,3 while in Sect\,4 the analyzed sample of X-ray sources is considered in 
the view of source population studies in the Magellanic Clouds and nearby galaxies 
in general.

\begin{acknowledgements}
The XMM-Newton project is supported by the Bundesministerium f\"ur Bildung und
For\-schung / Deutsches Zentrum f\"ur Luft- und Raumfahrt (BMBF / DLR), the
Max-Planck-Gesellschaft and the Heidenhain-Stif\-tung. The USNO A2.0 catalogue was
produced by the US Naval Observatory.
\end{acknowledgements}

\bibliographystyle{apj}
\bibliography{mcs,general,cv,myrefereed,myunrefereed,mytechnical}

\end{document}